# 4D Liquid-phase Electron Microscopy of Ferritin by Brownian Single Particle Analysis.

*Gabriele Marchello*[1,2,3,‡], *Cesare De Pace*[1,2,3,‡], *Neil Wilkinson*[5], *Lorena Ruiz-Perez*[1,2,3,\*], *and Giuseppe Battaglia*[1,2,3,4,6,7,\*]
[1]*Physical Chemistry Chemical Physics Division Department of Chemistry,* [2]*Institute for the Physics of Living Systems,* [3]*EPSRC/JEOL Centre for Liquid Phase Electron Microscopy,* [4]*Department of Chemical Engineering, University College London, London, UK.* [5]*Gatan UK, 25 Nuffield Way, Abingdon OX14 1RL, UK.* [6]*Institute for Bioengineering of Catalonia (IBEC), The Barcelona Institute for Science and Technology (BIST),* [7]*Institució Catalana de Recerca i Estudis Avançats (ICREA), Barcelona, Spain.*
[‡]*These authors have contributed equally*
\**Corresponding authors: Dr Lorena Ruiz-Perez* [l.ruiz-perez@ucl.ac.uk](mailto:l.ruiz-perez@ucl.ac.uk), *and Prof Giuseppe Battaglia* [g.battaglia@ucl.ac.uk](mailto:g.battaglia@ucl.ac.uk)

**Abstract**

Protein function and activity are a consequence of its three-dimensional structure. Single particle analysis of cryogenic electron micrographs has radically changed structural biology allowing atomic reconstruction of almost any type of proteins. While such an approach provides snapshots of three-dimensional structural information that can be correlated with function, the new frontier of protein structural biology is in the fourth dimension, time. Here we propose the use of liquid phase electron microscopy to expand structural biology into dynamic studies. We apply here single particle analysis algorithm to images of proteins in Brownian motion through time; thus, Brownian single particle analysis (BSPA). BSPA enables to reduce the acquisition time from hours, in cryo-EM, to seconds and achieve information on conformational changes, hydration dynamics, and effects of thermal fluctuations. Yielding all these previously neglected aspects, BSPA may lead to the verge of a new field: dynamic structural biology.

Electron microscopy (EM) is one of the most powerful techniques for structural determination at the nanoscale, with the ability to image matter down to the atomic level. The short wavelength associated with electrons is pivotal for achieving such a high resolution. However, electrons interact strongly with matter making imaging possible only under ultrahigh vacuum. Such a necessity limits EM imaging to solid-state or non-volatile samples. Samples in most common liquids or containing liquids- especially water- thus require special preparation techniques involving either controlled drying or cryogenic treatments with consequent artefacts. Such alterations become particularly critical for biological samples and soft materials whose meso- and nano- scale structure comprise water as a building block. Some of these limitations can be overcome using fast vitrification processes to solidify liquid samples (*1*), vitrified water is not liquid water and its structure, and hydrogen bond network are very different(*2*, *3*). For the last three decades, cryogenic electron microscopy (cryo-EM) has radically changed structural biology (*4–6*). Single particle analysis (*7*, *8*) in particular, is enabling atomic reconstruction of almost any type of proteins(*9*) with the tremendous consequence in drug discovery (*10*). While cryogenic and dry TEM provides snapshots of three-dimensional structural information critical to elucidate function, the new frontier of materials science and biology is in the fourth dimension, time(*11*).
The recent development of electron-transparent materials (*12*), (*13*) have paved the way to liquid-phase electron microscopy (LP-EM) (*14*). EM imaging of samples in liquid have been reported for electrochemical reactions (*15*, *16*), nanocrystal growth (*17*, *18*), labelled biostructures in biological cells (*19*, *20*), and tomography reconstruction of particles in the liquid state (*21*) among others. In LP-EM, the 3D reconstruction of particles in liquid has been mainly dominated by the analysis of inorganic particles (*21*), (*22*).
Here we exploited LP EM to image the dynamics of particles undergoing Brownian motion (*23*) to reconstruct their structures. Consequently, the technique here presented and discussed is named "Brownian single-particle analysis (BSPA)". In LP EM, the 3D structure of proteins has been studied entrapping them by functionalising the substrate (*24*), yet dynamics has never been involved in the analyses. We have investigated ferritin in its liquid state environment; ferritin is a biologically essential protein for cellular regulation of iron, already epitomised in LP EM (*24, 25*). Ferritin was imaged in its liquid native state, and the captured profiles were used to reconstruct the 3D density maps of the ferritin, through BSPA. This technique has its roots in single particle analysis (SPA), a state-of-the-art procedure dominating in 3D EM reconstruction. Conventionally, SPA processes images containing thousands of particles of the same nature showing random profiles. It requires samples in solution to be immobilised on a grid to preserve their structures under vacuum (*26*). Conversely, BSPA can process videos that may not contain thousands of particles, and monitor particle profiles alongside different frames, rather than across a single frame. Therefore, the orientations needed for the reconstruction are produced by recording the Brownian motion of the particles in solution. Time results to play a pivotal role in the reconstruction process, and a few seconds of videos can replace hours of samples preparation in cryo-EM.

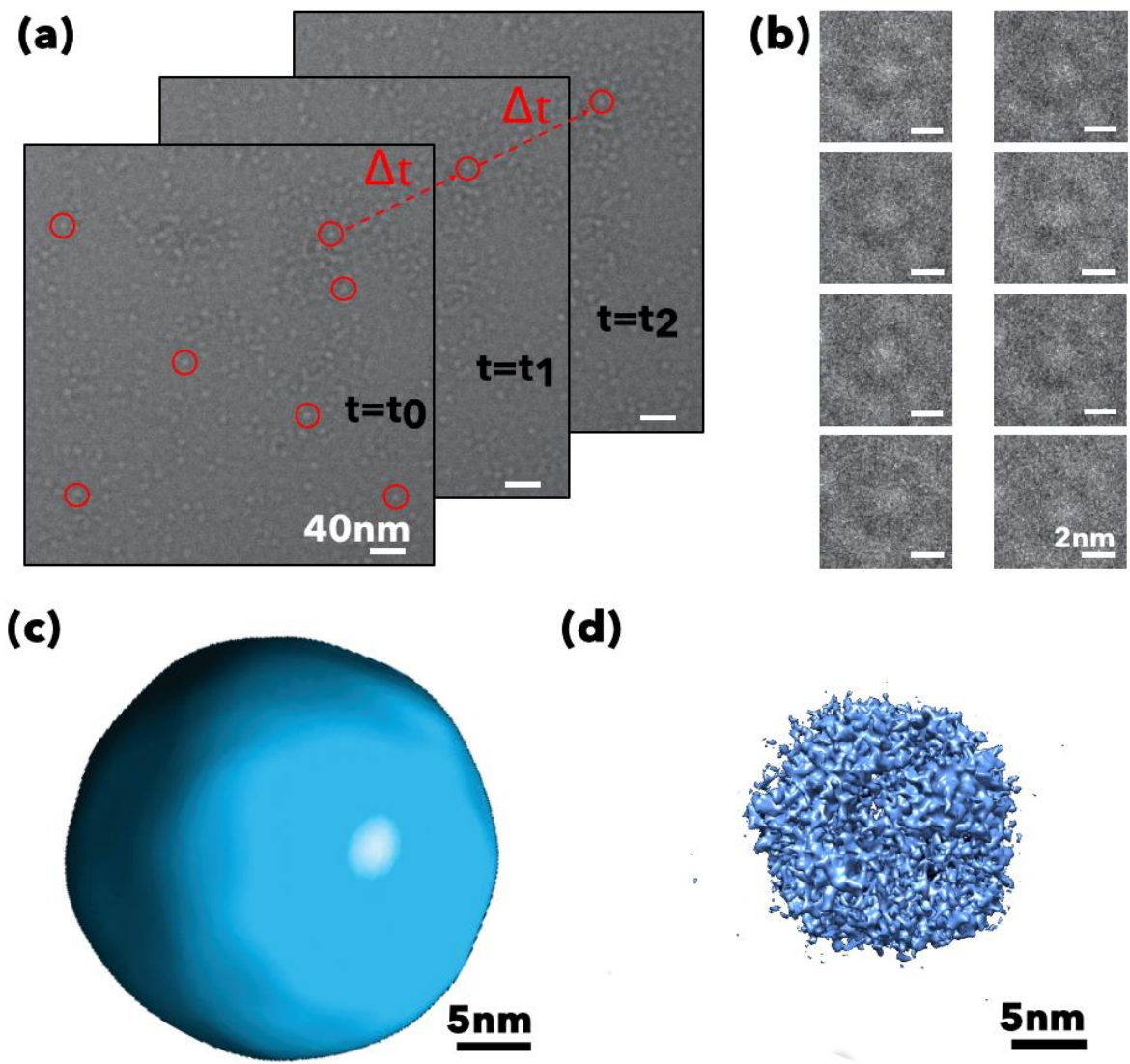

*Figure 1* – Schematic diagram showing the 3D reconstruction process of ferritin particles imaged in PBS solution using the Brownian single particle analysis (BSPA). ***a)*** *a single particle of ferritin being tracked across three consecutive frames and* ***b)*** *various profiles of the same ferritin particles with their associated angular orientations obtained from eight consecutive frames* ***c)*** *Initial coarse 3D density map reconstructed by Scipion from all the particle profiles extracted from a)* – ***d)*** *refined* **3D** *density map by means of a projection matching algorithm* (*28*).

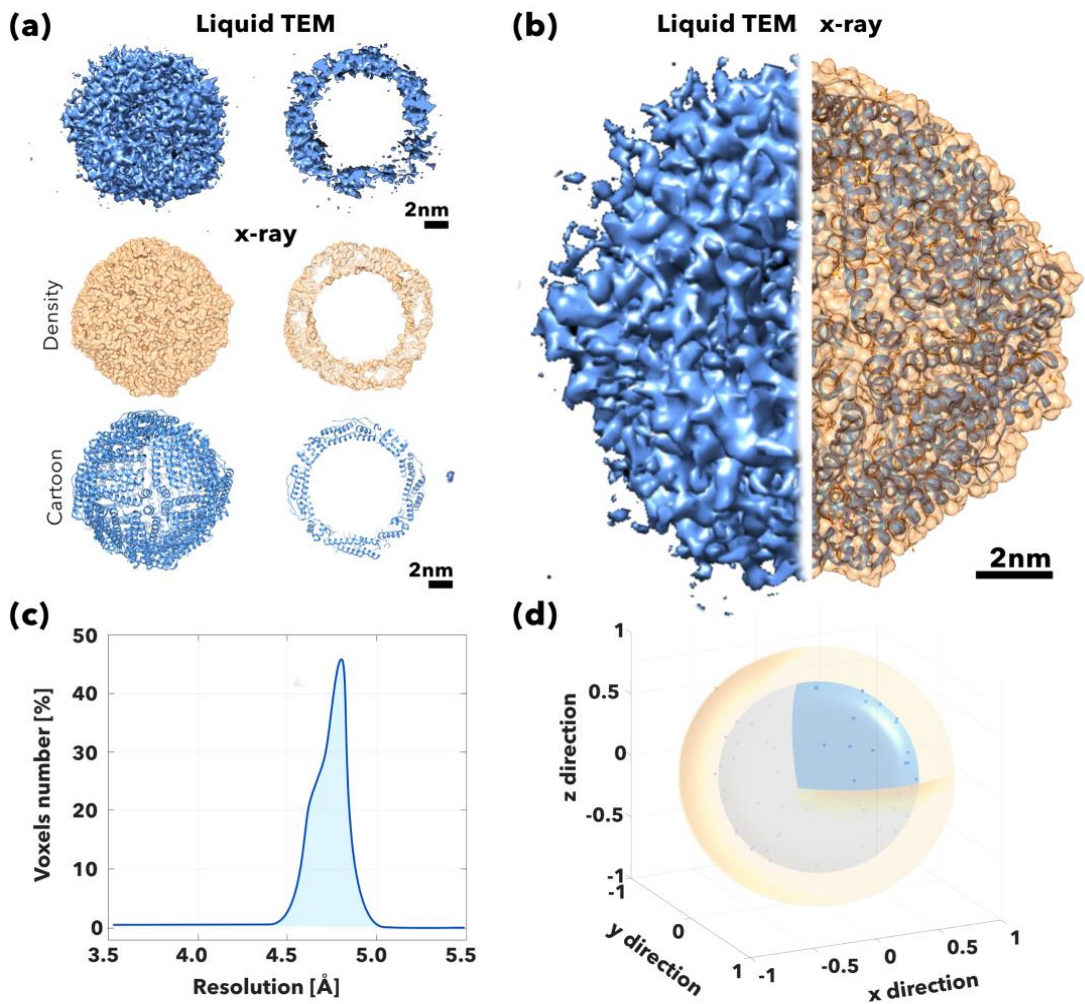

*Figure 2* – Comparison between liquid TEM and x-ray reconstructed 3D density map (in blue) of ferritin and the ground truth (in orange). *The density map of ferritin was while the 3D model was **(a)** 3D density map generated by the Brownian single particle analysis reconstruction technique and the model downloaded from the protein database (PDB) file:3F32 and reconstructed from X-ray diffraction. The models are displayed both as top view and the equatorial section with the x-ray reconstructed structure shown as density map and ribbon cartoon to highlight the protein folding. **(b)** Halves of the reconstructed 3D map and model of ferritin displayed together highlighting the similarities between both structures. The size is comparable, and the pattern of the helixes in the ground truth continues in the noisy pattern of the reconstructed density map. **(c)** Histogram showing the local resolution of the reconstructed map after being computed. The histogram displays the voxels number versus resolution values expressed in Angstrom. The number of voxels was normalised to facilitate further comparisons **d)**. Cross-correlation analysis between liquid TEM and x-ray models reported in a 3D graph with a blue marker in accordance with the map rotations. The distance between the markers and the origin of the plane represents the similarity of the projections. The orange sphere represents the perfect match between the maps and has unit radius while the blue is the average between the measured correlation and has radius 0.72.*

Ferritin in phosphate buffered saline (PBS) solution at 2.5mgml$^{-1}$ was imaged as described in the *Supporting Information*. Ferritin particles were recorded, displaying Brownian translational and rotational motion. Brownian rotation means that each particle dispersed in the liquid will show several faces under the TEM at a frequency inversely proportional to the cubic radius and which for the ferritin here corresponds to approximately 10$^4$Hz. The video employed for the BSPA reconstruction technique displayed a few hundreds of ferritin particles in each frame and was recorded for ten seconds at 10Hz, generating a hundred frames. Over 5000 particles were extracted from the first fifty recorded frames and used to reconstruct the density maps. Firstly, the stack of two-dimensional (2D) micrographs was pre-processed to reduce the noise. The core of the reconstruction algorithm was performed with Scipion, a freeware designed to perform SPA 3D reconstruction, that was extended and adapted to our needs (*27*). The imaged particles were identified across all the frames, extracted and collected together to generate a 3D density map of the ferritin protein. ***Fig. 1a*** shows a single particle of ferritin being tracked across three consecutive frames. An example of the variety of particle profiles acquired over time is shown in ***Fig. 1b*** for eight consecutive frames. The BSPA method can be considered a reference-free technique because it does not use any *a priori* information about the imaged particles. A coarse 3D density map was initially reconstructed and shown in ***Fig. 1c*** and the map was subsequently refined by applying the projection matching algorithm to produce an unbiased, high-fidelity density map, as shown in ***Fig. 1d***.

A detailed description of the analysis method is provided in Supporting Information – 1.5 Reconstruction. It is worth mentioning that the generated coarse and refined density maps, shown in ***Figs. 1c-d*** respectively seemed to vary in diameter. A plausible explanation for the coarse density map displaying a diameter of almost twice the diameter of the refined map could be the inclusion of a hydration layer surrounding each particle in the coarse map. This hydration layer would indeed generate 'swollen maps' and account for the liquid nature of the sample.

In order to validate the Brownian single particle analysis method, the generated 3D density map of ferritin was compared to the well-known structure present in the protein data bank (PDB file: 3F32) (29). A graphical result of the final 3D reconstructed density step is shown in ***Fig.2.a*** and compared to both the PDB model represented as a density map or as cartoon to highlight the different secondary structures. The BSPA 3D map and ground truth structures showed a match in shape, size and dimension, measuring both 12nm and circa 2nm in their outer diameter and shell thickness, respectively (***Fig.2b***). The reconstructed density map was further validated by using the ResMap algorithm to measure the local resolution (30). This procedure determined both mean and median resolutions of 4.74Å and 4.80Å (***Fig.2c***). We further compare the two models quantitively by aligning the two density maps are and fixing this as the origin [0 0 0]. The maps were rotated along the three axes of finite and equal increments, and the consequent 2D projections were compared by normalised cross-correlation. The resulting similarity values were expressed as the distance from the origin of a 3D space, oriented according to the rotation of the maps (***Fig. 3b***). The perfect match between the maps was overlapped to the results space and was represented as an orange sphere of unit radius. This sphere was plotted as reference in the 3D space. All the results of the comparison resulted to be contained inside a sphere (blue) of radius 0.725, implying a mean similarity value equal to $(72.5\pm0.7)\%$.

The BSPA technique did not exhaust its task of reconstructing 3D density maps of ferritin, but it was crucial for analysing the evolution of the recorded structures over time. The same video used for the reconstruction was segmented into five consecutive, not-overlapping intervals, containing ten frames each. These five one second long video segments were processed, and the associated 3D density maps were reconstructed. Each video segment contained about 1000 different particles. The apparent low number of particles could lead to low-quality density maps, however, both coarse and refined maps (***Fig. 3a***) yield useful information about the evolution of the particles in time. The generated density maps from each video segment, displayed a variation on the contour and morphology of the particles shell over time. Although the particles being the same type should lead to five identical 3D density maps, different configurations of the maps surface and shell outer contour were obtained. Such differences observed over one second time intervals could be explained by thermal fluctuations effects deforming the particles soft surface or changing the hydration shell. It has become clear that long observation time increased the number of particle profiles; however, the longer exposure to the electron beam significantly damages the particles. The ResMap algorithm measured respectively resolution mean values of 5.42Å, 5.39Å, 5.39Å, 5.37Å, 5.37Å; and median values of 5.5Å, 5.4 Å, 5.4 Å, 5.4 Å, 5.4 Å (***Fig. 3b***). The difference in the resolution values between the five density maps from ***Fig. 3*** and the long-exposure density map from ***Fig. 2*** accounted for the difference in the number of particles involved in the reconstructions.

Consequently, the quality and resolution of the refined density maps resulted in being inversely proportional to the sample exposure time to the electron beam (***Fig. 3c***). The analysis of the resolution obtained from these five time-varying maps leads to interesting considerations. The number of missing patches on the density maps seemed to increase over, while the resolution of the density maps remained constant.

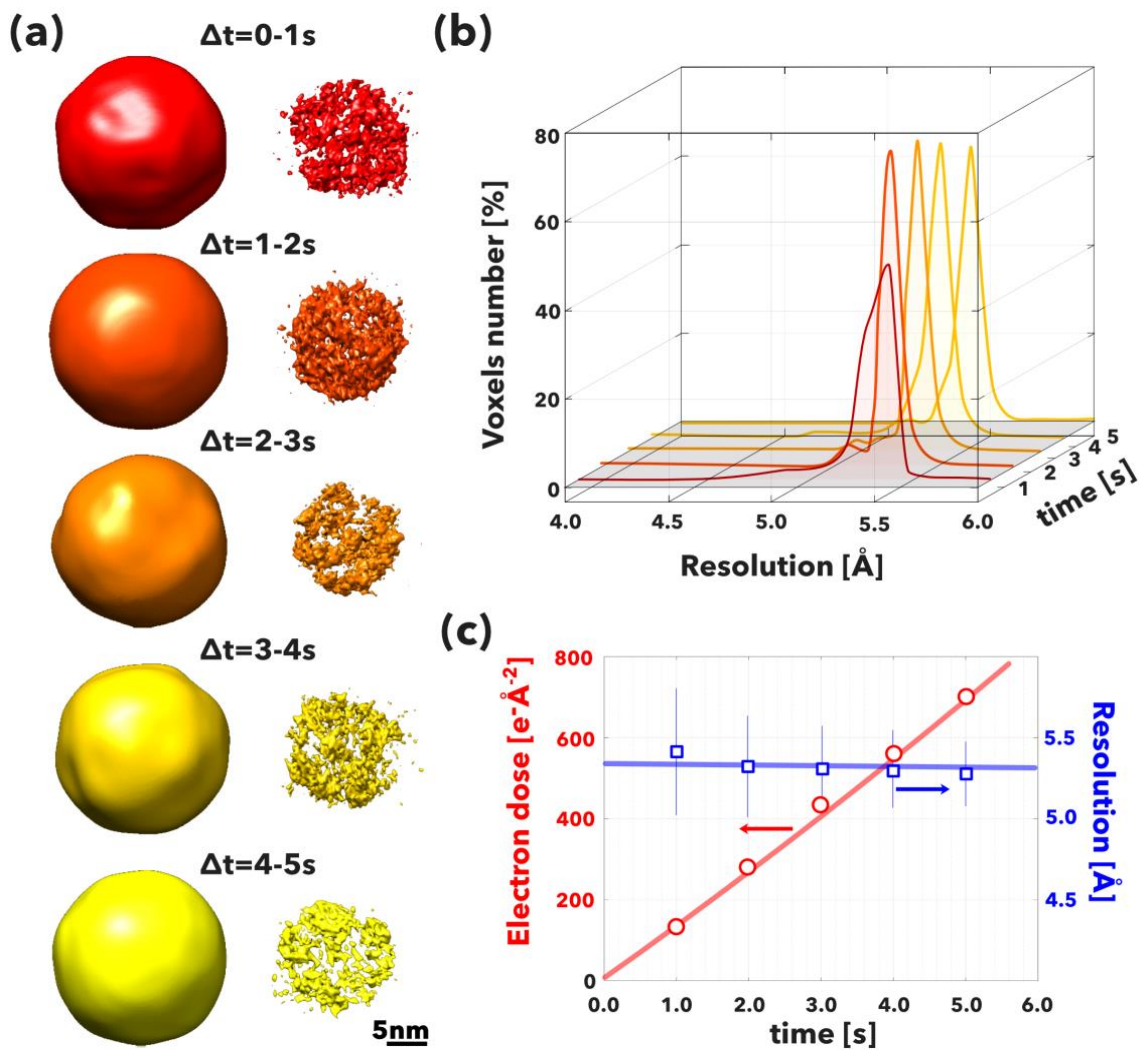

*Figure 3* – Dynamic reconstruction of ferritin as a function of time *(a) Schematic representation showing the temporal evolution of the density map reconstruction process. The algorithm extracted about 1000 particles profile from each sub-video, generating five different density maps shown as course and refine the model. (b) The histograms generated from the measurements of the local resolution of each density map performed through ResMap for each temporal segment. (c) The electron dose the particles were exposed to expressed for each 3D density map and plotted alongside the mean spatial resolution.*

In conclusion, the BSPA technique presented here proposes an entirely new way to reconstruct 3D density maps of protein in liquid solution avoiding possible artefact arising from the vitrification process. The time factor becomes a fundamental element to the reconstruction process, reducing the acquisition time required to collect enough particles from hours in cyroEM to seconds. Most notably, BSPA allows extending atom-level structural biology to dynamic structural biology shedding light on conformational changes, hydration dynamics or merely evaluating the effect of thermal fluctuations. We believe that our findings pave the way to a completely new frontier in structural biology expanding functional studies to dynamics where we can sample proteins within a three-dimensional plus time i.e. 4D space. Such a space will allow us to understand conformational changes, allosteric interaction, folding-unfolding as well as to bring the so far neglected role of water in protein structure.

**Acknowledgements:** The authors would like to thank the EPSRC (EP/N010906/1), Jeol, DENSsolutions and Gatan for sponsoring the work, LRP salary, CDP and GM studentships. GB thanks to the ERC (Grant CheSSTaG 769798) and the EPSRC (EP/N026322/1) for covering part of his salary.

# Supporting information

**Methodology**

Liquid cell preparation and assembly. The holder used for imaging ferritin in liquid state was the *Ocean* liquid holder manufactured by *DENSsolution*. The ferritin in solution was sandwiched in a liquid cell formed by two chips made of silicon nitride ($Si_xN_y$), with 50nm thick rectangular observation windows at the centre measuring 20μm x 200μm. Silicon nitride prevents the solution to leak from the closed cell and does not obstruct the path of the electrons through the sample as it is electron transparent. The liquid cell is comprised by a spacer-chip, placed at the bottom and a spacer-less chip placed at the top. 200nm spacer chips were used for the reported investigation. Although the liquid thickness is dictated by the spacer used in the liquid cell, the observation windows of the chips tend to experience some bulging, mainly at their centre point. This bulging effect is due to the differential pressure encountered by the holder once it is inserted in the microscope. The bulging phenomenon of the windows consequently adds some extra thickness to the observation chamber, typically up to 150nm in excess per chip, to the 200nm thickness provided by the spacer. The bulging effect is minimum however at the corners of the rectangular windows. Thus ferritin imaging was performed by the areas closed to the corners of the $Si_xN_y$ window. Nonetheless, the thickness of the liquid layer resides in the nanometre range, allowing for good electron penetration and sample contrast.

Equine spleen apoferritin was purchased from Merck (previously Sigma-Aldrich) and diluted with phosphate buffer solution (PBS) at a concentration of 2mg/ml.

First, the chips were rinsed in HPLC-graded acetone followed by isopropanol for five minutes each, to eliminate their protective layer made of poly(methyl methacrylate) (PMMA). Then, the chips were plasma cleaned for thirteen min to increase their hydrophilicity. Soon after, the chip with the spacer was fitted at the bottom of the liquid cell and where 1.5μL of ferritin solution was deposited on the window. It was important not to let the ferritin solution dry at this stage as the film thickness of dried ferritin solution cannot be easily controlled and impedes good imaging conditions. The chip without spacer was then positioned on top of the liquid layer closing the liquid cell. The liquid holder was sealed, and 300μL of the ferritin solution at 2mg/mL was flushed in at 20μL/min with a peristaltic pump. 300uL of solution volume ensured that the holder tubing system and cell were filled with ferritin solution. We waited five minutes after collecting ferritin solution from the outlet tube (Fig.1 *SI*) in order to guarantee that convection effects from the flowing process were not affecting the Brownian movement of ferritin in solution. The liquid holder was then introduced into the microscope. Fluid dynamics simulation performed on the liquid holder are reported in section *2.1 Fluid dynamics simulation*.

LP EM imaging procedure. The experiments were performed using a JEOL JEM-2200FS TEM microscope equipped with a field emission gun (FEG) at 200 kV, and an in-column Omega filter. The camera used was the direct detection device (DDD) *in-situ* K2-IS camera from Gatan. The ultra-high sensitivity of the K2 allows low-dose imaging modes limiting considerably electron dose damage and facilitating high spatial (3838x3710 pixels) and temporal (up to 1600 fps in *in-situ* mode) resolution.

The microscope was used in transmission electron microscopy mode. In order to limit the beam dose on the specimen, images were collected at the minimum spot size (#5) with a small condenser lens aperture (CLA #3). The K2 can be used in two recording modes, linear and counted. The former is fast, but it generates the output image by capturing the integrated signal from each pixel, as a standard charge coupled device (CCD). Counted mode takes only the pixel with the highest intensity in a neighbourhood of four pixels. For our investigations dose fractionation videos were recorded in linear mode. Further details on the quality of the videos are provided in section *2.2 Video analysis*. The electron flux used to record the videos was 13.9 e-/Å².

Underfocus of circa 10μm was needed to improve the contrast of ferritin as phase plates were not present in our imaging system. Every image was recorded in the format of 4-byte grayscale and required the full size of the detector. However, the images were binned by a factor 2 on both dimensions, resizing images down to 1919x1855 pixels. This process not only increased the signal-to-noise ratio (SNR) of each processed frame, but also prevented the memory of the processing machine to saturate due to the high volume of data

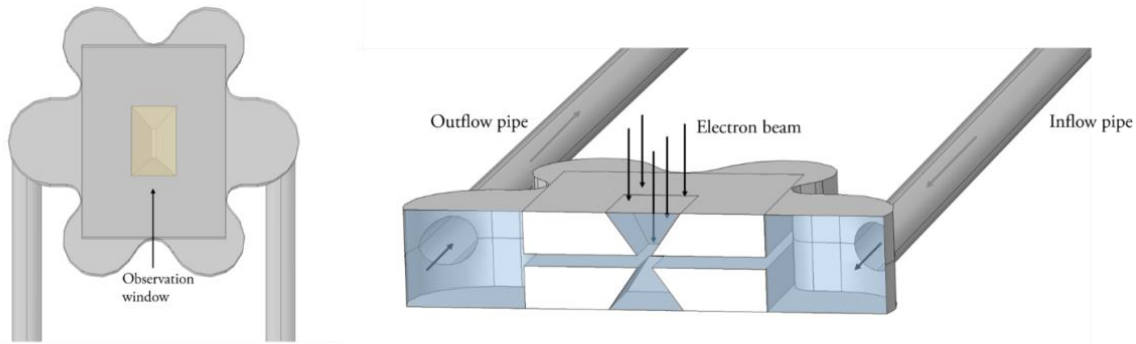

*Figure S1 -* Schematic representation of the liquid cell used for liquid-phase electron microscopy (LP EM) imaging. The liquid holder was manufactured by DENSsolutions and reconstructed in 3D by using COMSOL Multiphysics. *(a)* The top-view image highlights the flower-shaped tip of the holder, which is connected to two pipes placed on the same plane and appointed for flowing liquid in and out of the holder. The holder hosts on the inside two microchips, with observation windows placed at the centre (in yellow). *(b)* The cross section of the holder. The sample is pumped in the chamber, defined by a variable gap, i.e 200nm that creates a microfluidic channel to enable inflow analysis and keep the sample hydrated. The electron beam is set perpendicular to the holder and passes through the electron-transparent windows, before being collected by the detector of the camera underneath the holder.

Video Pre-processing. The videos were pre-processed using an in-house MATLAB script [1], with the aim to adapt all the recorded frames to best fit the reconstruction process requirements. Firstly the images were separated from the metadata in the .dm3 files, which were secondary to the reconstruction process and did not require further alterations. The image data resulted to be significantly affected by noise and needed to be reduced as much as possible, without losing details. Therefore, images were first filtered by applying a median filter and then normalised with respect to the maximum intensity pixel value. The median filter was employed to remove the so-called salt-and-pepper noise corrupting the images; more precisely, it removed the very white and very black noisy pixels [2]. Normalising the value of the pixels after the filtering process increased the contrast of the images. After pre-processing all the images, they were exported into mrc format files, as required by *Scipion*.

Ferritin Reconstruction. *Scipion* is a freeware designed for single particle analysis (SPA), which offers a modular approach to solve the reconstruction problem, integrating dozens of tools fitting the needs of the users [3]. Consequently, several different workflows can be generated, but, in the present investigation, only a single standardised solution is presented and discussed. This workflow was proven to be general and efficient for many different sets of micrographs, even if the first steps may differ with respect to frame rate value and SNR. The first step was to import the micrographs, together with some parametric and descriptor values of the imaging process (*Table 1*), such as pixel size expressed in angstrom per pixel (Å/px), microscope voltage in kiloelectronvolt (k eV), spherical aberration in millimetres (mm) and magnification rate. These parameters proved to be necessary to recreate an in-scale high-fidelity 3D density map. Most of these values were associated to the microscope and, therefore, were considered constant, while the magnification varied according to the settings of the imaging process (*Table 1*).

*Table 1 - PARAMETRIC AND DESCRIPTOR VALUES OF THE IMAGING PROCESS*

|  | *Video1* |
| --- | --- |
| *Pixel size (Å/px)* | 1.2 |
| *Microscope voltage (keV)* | 200 |
| *Magnification rate* | 10,000 |
| *Number of frames* | 50 |
| *Original frame dimension (px)* | 3838x3710 |
| *Binning factor* | 2 |
| *Frame rate (fps)* | 10 |
| *Observation time (s)* | 5 |

The second step consisted on correcting the contrast transfer function (CTF) [4] of all the images, a function modelling the distortion introduced by the microscope during the imaging process. The estimation of the CTF and the following correction step of the workflow aimed to restore the original and undistorted object from all the micrographs. In the workflow discussed in this paper, these operations were performed by using the tool CTFFIND4, developed by Gregorieff [4]. The following third step was to identify the particles in the frames, with the final aim to create a gallery collecting all the 2D particle profiles. This tool was implemented by EMAN2 [5], an image processing set of functions widely used for SPA and 3D reconstruction, and fully integrated in Scipion. The particle identification step could be performed either manually or automatically, but in the present reconstruction a supervised training algorithm was set up (*Fig. 2*). This selection mode speeds the process and, at the same time, reduces the number of false positives; patches (*i.e.*, windows of pixels) of the images erroneously identified as particles. The supervised training algorithm required the users to manually select (black boxes in Fig. 2a *SI*) a few particles in the first frame to train the software creating a template of the ferritin particle (*Fig. 2b SI*) to identify other ferritin particles.

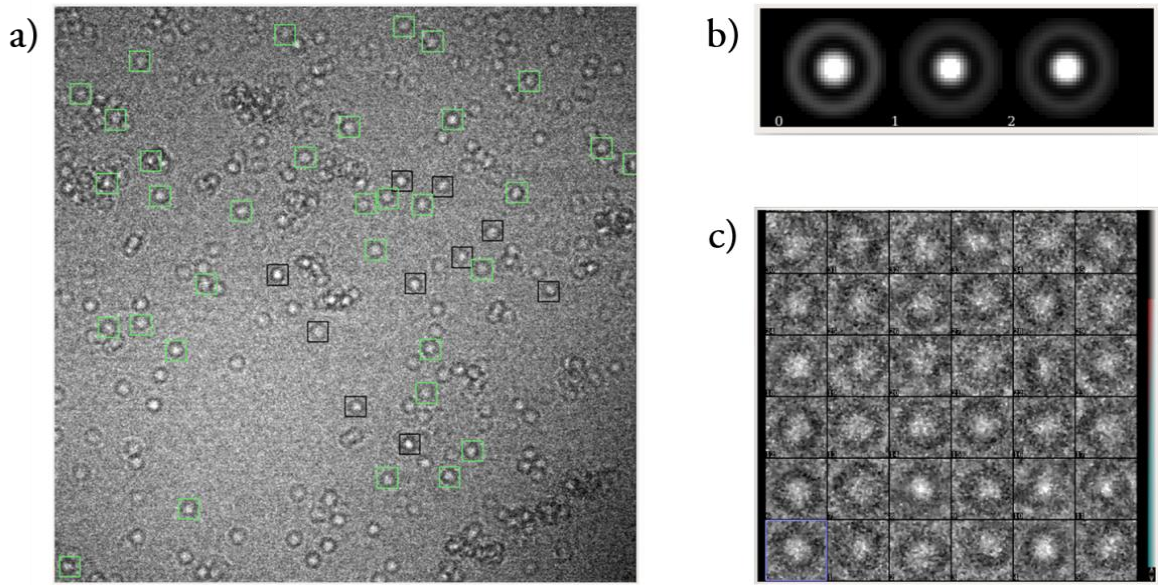

**Figure S2 –** Schematic representation of the ferritin particle selection process. *(a) The black boxes represent the ferritin particles selected by the user, which are transformed into a template. (b) The template is then used to recognise and identify all the occurrences in the rest of the frame (in green). (c) The particles automatically selected are saved into a gallery. Lastly, the so-defined particle identification scheme is applied to the rest of the frames comprising the videos.*

Then, the users have to refine the automated particle selections of the software, manually discarding wrong results. Once EMAN2 was trained, the software applied the results of the training algorithm to the remaining frames, automatically identifying the 2D coordinates of every particle in green as seen in Fig. 2a *SI)*.

Table 2 - PARAMETERS RELATED TO THE PARTICLE SELECTION STEP

| | |
|---|---|
| **Training frames** | *1* |
| **Particles picked manually** | ≥5 |
| **Particles extracted** | 5256 |
| **Window size (px)** | 120 |
| **Particle size (px)** | 100 |

The software also offered a further possibility to process the selected particles before exporting them into the gallery. Removing the background and inverting the contrast were the most important functions, amongst the many built-in functions. The former identified as background all the pixels in the patches outside a circle, the radius of which is defined by the user. Consequently, pixels carrying useless or misleading information were cut off from the reconstruction algorithm. Inverting the contrast was necessary to adapt the particles to the following steps, since most of the algorithms included in Scipion work with white particles on a dark background. Conventionally, SPA reconstruction techniques can be categorised into two big categories, depending on the need for an *a priori* 3D model used as reference [6]. All the reference-based algorithms assigned an angular orientation to every two-dimensional (2D) projection by minimising the distance with the model. Consequently, the reference model significantly biased the final reconstruction. However, the absence of the so-called ground truth made the reconstruction of the final volume difficult. RELION [7] overcame this limitation reconstructing an initial 3D model by using a stochastic gradient descent (SGD) algorithm [8]. SGD is an iterative algorithm that tries to find the reconstructed volume that is best described by the set of angular orientation assigned to the 2D image subsets. The representation adopted in the present work to describe the angular orientation of a body is roll-pitch-yaw (RPY), which describes consecutive rotations of the body around the three axes of the fixed reference system [9]. In order to facilitate the 3D reconstruction, RELION allowed the user to define the direction of an axis of symmetry. However, no axes of symmetry were set with the aim to remove any possible source of bias. This process generated an initial, coarse map working as reference, which was further processed to produce the final, high-fidelity 3D map. This refinement step was performed by Xmipp through projection matching, an algorithm based on 3D maximum likelihood multireference refinement (or ML3D classification) [10]. This method consisted in finding the set of angular displacements that maximised the likelihood function applied to the projection of the coarse map and the 2D micrographs. Although Xmipp offered the opportunity to select the reconstruction method amongst Fourier space interpolation, algebraic reconstruction technique (ART) and the weight back projection (wbp) methods, in this workflow the Fourier technique was chosen. In order to validate the reconstruction process, the resolution of the final 3D map was measured through ResMap [11].

Correlation between the 3D reconstructed density map and ground truth model. After the reconstruction of the 3D density map of ferritin was finalised a validation step was implemented to measure the similarity between the reconstructed map and the density map obtained *via* cryo-EM and registered in the PDB library [12]. The maps were aligned by hand and then a fixed reference frame was inserted in the middle of both maps (*Fig 3a*). Both maps were rotated along the three axes by a constant increment of 30°, in order to cover most of the angular orientations. After each rotation, the 2D projections of the maps were stored and analysed by using an in-house MATLAB script. First, the images were transformed into grayscale to avoid any bias due to the different colours. Then, this script computed the similarity of each set of projections through normalised cross-correlation [13]. These similarity values were expressed as distance from the origin of a 3D space and oriented according to the rotation of the maps (*Fig 3b*). The perfect match (i.e 100%) between the maps was plotted together with the cross-correlation results in the 3D graph and represented as a pink sphere of unit radius. This sphere was used as reference in the 3D graph. All the results from the maps comparison resulted to be contained within the pink sphere. This implied that the match between our reconstructed density map and the ground truth model of ferritin along the different angular orientations was not perfect. The similarity between the maps was computed as a local feature, however a global value was measured by averaging all the results. This process led to a global mean similarity value equal to $(72.5\pm0.7)\%$ between the reconstructed 3D density map of ferritin and the ground truth model. This method resulted to be significantly biased by the choice of the first alignment, which was then chosen as reference angular orientation [0,0,0] of the maps. However, this bias can be used as an interesting tool to run a quantitative analysis on the shape of the reconstructed maps.

Performance analysis. The whole workflow was executed on a machine running Ubuntu Linux 16.04 64-bit operative system with the following hardware specifications: two central processing units (CPU) Intel Xeon Gold 5118 2.3GHz with 12 cores each, 128GB (8x16GB) 2666MHz DDR4 RAM, 512GB class 20 solid state drive (SSD) and dual SLI NVIDIA Quadro P5000 16GB as GPUs. Table 4 reports the processing time values associated to the single steps of the workflow with respect to the processed videos.

*Table 3 - PROCESSING TIME VALUES RELATED TO THE STEPS OF THE WORKFLOW*

| **MATLAB pre-processing (per frame)** | 00h 00' 08" |
|---|---|
| **Pre-process micrographs** | 00h 04' 29" |
| **CTF correction** | 00h 02' 48" |
| **Particles picking** | 00h 03' 45" |
| **Particles extraction** | 00h 00' 55" |
| **3D initial map** | 00h 05' 02" |
| **3D refinement** | 02h 38' 03" |
| **Local resolution** | 00h 06' 54" |

## Supplementary discussion

<u>Fluid dynamics simulation.</u> Section *1.1 Liquid cell preparation and assembly* shows how demanding this technique can be, and it is not surprising that the first attempts to image samples in liquid state failed due the high complexity of the system. Therefore, a fluid dynamics simulation was performed by using *COMSOL Multiphysics* in order to understand the functioning of the liquid cell in *in-flow* conditions and obtain the best performance from the imaging process. The liquid holder was modelled in *COMSOL* respecting the actual dimensions of the device, the values of which will not be reported in this paper not to violate the confidentiality agreement signed between the authors and the manufacturing company. However, the simplified geometry of the holder is shown in *Fig. 1*. The tip of the holder is flower-shaped and is connected to two in plane parallel pipes, an inlet and outlet pipe for flowing in and out the solution respectively. The holder hosts on the inside two SiN chips separated by a variable gap, creating a microfluidic channel to enable inflow analysis and keep the sample hydrated. The bulging effect on the observation windows occurring when imaging the sample and mentioned in section 1.1 of *SI* has been excluded from the simulation. This is because the simulation has been performed at atmospheric pressure conditions i.e. when the sample is flowed in the liquid holder outside the microscope. Thus the windows are treated as flat, rigid surfaces trapping the solution on the inside. Input and output ports were set to the end of the pipes, guaranteeing good fluid circulation. At the beginning of the simulation the chamber was modelled as empty, both chips were removed, and an uncompressible Newtonian fluid was pumped in with an initial velocity of 5µL/min. The analysis was performed at steady state, imposing laminar flow, planar symmetry and no slip condition on the walls of the holder. The velocity of the fluid in these conditions was analysed and the results are reported in *Fig. 3*. This first preliminary simulation was mainly performed to validate the goodness of the model, and especially to analyse the trajectory of the flow. The first simulation confirmed that the fluid linearly went from the inlet to the outlet pipes, slowly expanding in the whole volume of the chamber. The results showed that the trajectory of the injected particles resulted to be a combination of both convection and Brownian motion experienced by nanoparticles in liquid. Then, the complexity of the system was increased by adding the two chips inside the flower-shaped tip, separated by a 200nm spacer. The inclusion of the chips significantly influenced the liquid flow inside the chamber, modifying the liquid trajectory and narrowing it down to the small microfluidic channel between the chips. Consequently, most of the fluid went into the petals of the structure, generating a turbulent flow. Furthermore, the volume between the two chips was filled with only a small portion of the incoming fluid, the motion of which was hardly oriented by the effect of the pump (*Fig. 2b*). In a first approximation model, these results can be extended to the particles in solution pumped in the inflow port; mirroring the actual experimental conditions. As a consequence, most of the particles would not flow over the observation window, but instead be trapped in the flower petal reseravoir. These findings strongly suggest that parameters such as sample concentration and velocity at which the sample is injected into the holder have to be carefully chosen, in order to maximise the number of the particles in the viewing window area..

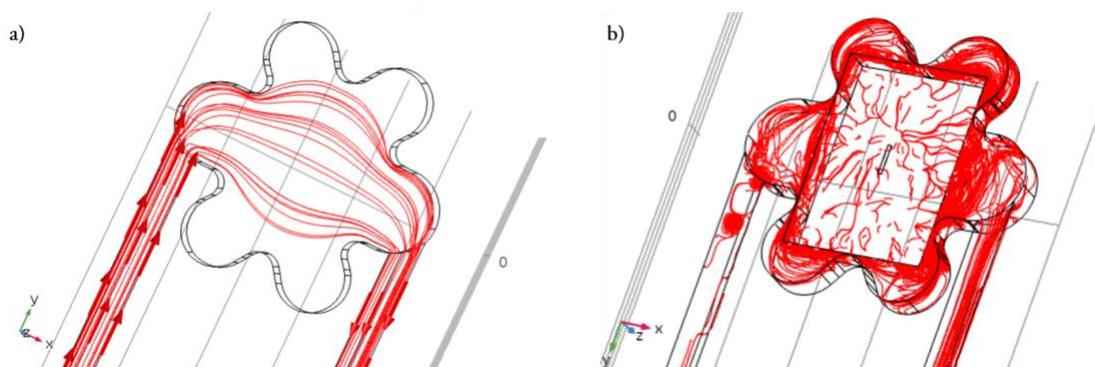

**Figure S3 -** Schematic representation of the fluid dynamics simulation performed on the Ocean holder. The liquid holder was modelled in 3D using COMSOL Multiphysics. *(a) Preliminary simulation involving a simplified version of the holder, modelled without the SiN chips. A Newtonian fluid was pumped in the inflow pipe (on the left) and out from the outflow pipe (on the*

*right). The flow expands towards the flower shape reservoir perpendicular to the liquid main trajectory while flowing through the chamber.(b) The first simulation was then modified by introducing the two chips inside the tip of the holder, in order to replicate the actual experimental condition. In this case, most of the fluid moved to the flower petal reservoir, with only a small a volume flowing between the chips. Both simulations were run applying laminar flow, planar symmetry and no slip conditions; the analysis was then performed at steady state. In this figure, only the lower half of the model is depicted, in order to facilitate the analysis.*

Video analysis. The reconstruction process of the 3D density map of ferritin in solution recorded by the K2 in dose fractionation video mode has been analysed and discussed above. However, not all the obtained videos could be used as input to the workflow. Since the sample is in liquid state, it was impossible to record twice the same video as the particles moved with time. Consequently, some parameters of the imaging process could vary, affecting the quality of the videos, even if the recording process was standardised. In this section, 12 different videos recorded during the same session are analysed. *Table 4* contains information about the recording mode, the frame rate, the number of frames and the observation time for each video.

*TABLE 4 – METADATA EXTRACTED FROM THE VIDEOS OF FERRITIN IN SOLUTION*

| Rec. mode | File name | Frame rate (fps) | Num. frames | Obs. Time (s) |
|---|---|---|---|---|
| Counted | Ferr_200nm_17 | 10 | 100 | 10 |
| | Ferr_200nm_18 | 20 | 200 | 10 |
| | Ferr_200nm_20 | 40 | 200 | 5 |
| Counted | Ferr_200nm_22 | 40 | 200 | 5 |
| | Ferr_200nm_23 | 20 | 200 | 10 |
| | Ferr_200nm_24 | 10 | 100 | 10 |
| Linear | Ferr_200nm_25 | 10 | 100 | 10 |
| | Ferr_200nm_26 | 20 | 200 | 10 |
| | Ferr_200nm_28 | 40 | 200 | 5 |
| Linear | Ferr_200nm_30 | 40 | 200 | 5 |
| | Ferr_200nm_31 | 20 | 200 | 10 |
| | Ferr_200nm_32 | 10 | 100 | 10 |

The recording mode has been widely discussed in section *1.2 LP EM Imaging procedure*, therefore, in this section other additional features are discussed. The file name of the videos is self-explanatory, as it contains the type of sample in solution, the size of the spacer between the chips in the holder and an identification progressive index. The frame rate value was varied between a minimum of 10fps and a maximum of 40fps, in order to improve the temporal resolution and record as many consecutive movements of the particles as possible. Therefore, the higher the frame rate value, the higher the number of orientations captured by the detector in a limited observation time. However, a very high frame rate value does not guarantee a high-quality reconstructed model, due to the low fraction of electrons collected by the detector and the consequent low contrast. Therefore, some of the frames recorded with a high frame rate were discarded, due to the poor quality of the image. 20fps and 40fps videos contained more frames and were recorded for a shorter observation time than the 10fps videos.

In order to select the video to use for reconstructing the density map of the sample, the quality of each video was analysed. However, the goodness of an image is a value very difficult to quantify and unfortunately, there are no objective ways to measure it. Nevertheless, an estimation can be computed by applying the peak signal-to-noise ratio (PSNR) proposed by Gonzalez [14]. This method estimates the quality of the image by measuring the squared difference between the noisy image and the original, "clean" image, and was proven to be correct for uniformly distributed noise (e.g. Gaussian). However, the imaging process associated to the LP EM produces images heavily corrupted as it adds the component due to the liquid between sample and detector to the conventional TEM noise [15]. Therefore, Gonzalez's method is used to get a qualitative analysis of the goodness of the images registered varying the frame rate values. A second element adding complexity to this analysis is the need of the ground truth to estimate the PSNR, a missing piece of the puzzle. Therefore, in order to "create" the clean image, the progressive image denoising (PID) algorithm [16] was chosen as denoising algorithm. The selection of the PID algorithm relies on its ability to produce enhanced results, without generating artefacts. However, this algorithm is very CPU demanding and time consuming; it takes about 30 minutes to process a single frame made of 1919x1855 pixels. Consequently, only a subset of three frames for every video has been selected, these frames were selected at the very beginning, in the middle and at the very end (*Fig. 3*) of each video.

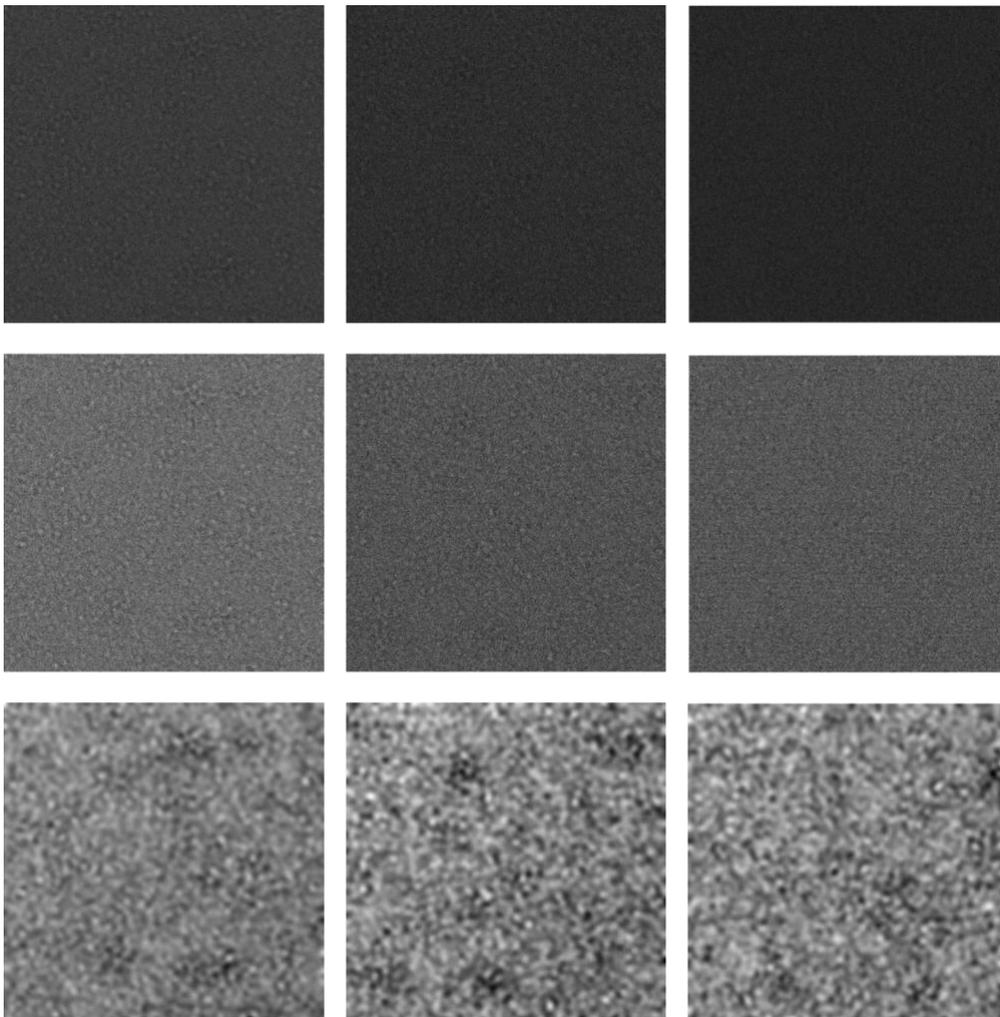

*Figure 4* – Schematic representation of the frames used to perform the PSNR analysis. *The columns of this matrix disposition represent the different frame rate values: from left to right, 10fps, 20fps and 40fps. The rows contain the different element of the analysis: from top to bottom, the original frame, the median filtered version of the original image and the reconstructed ground truth. The noiseless image at the bottom was obtained by applying the PID algorithm to the original image.*

The PSNR analysis has been run on the original frames, as discussed so far, and on median filtered images in order to validate the results of the algorithms. These analyses are reported in *Fig. S4*.

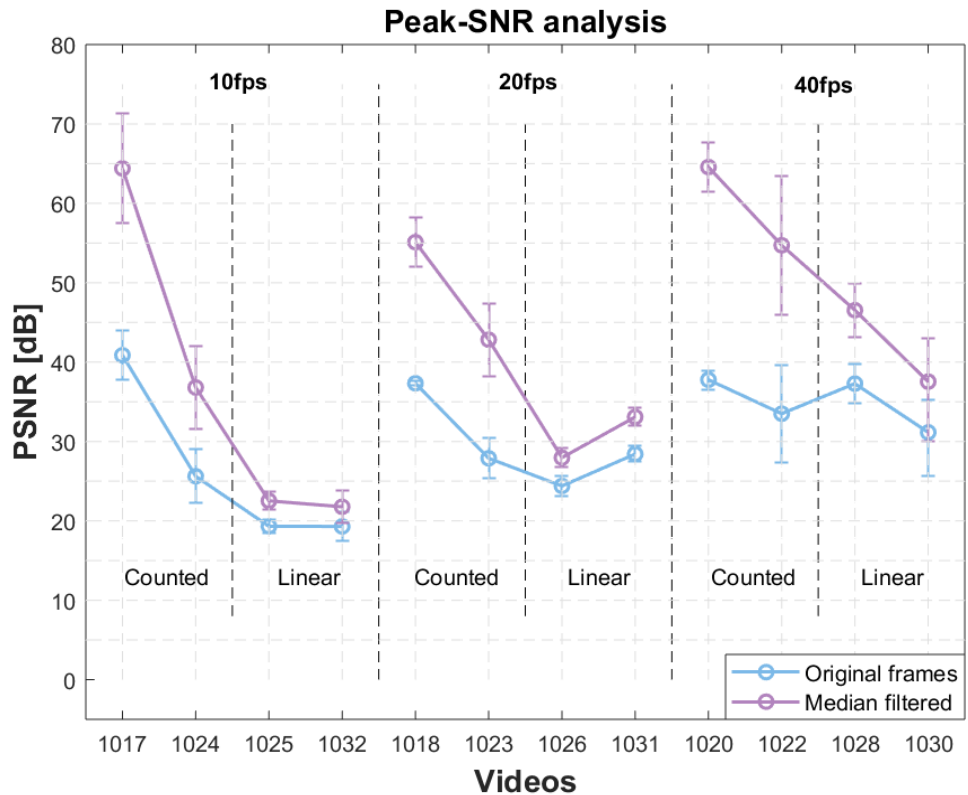

***Figure S5*** *– Graphs summarising the results obtained from the analysis of image quality, taking into account original videos (in blue) and the videos processed through the median filter (in purple). The PSNR of each video (reported on the x-axis and labelled with the identification number) is computed by using the Gonzalez method and expressed in decibel (dB). The videos were divided according to the frame rate value and each category was allocated into two subgroups according to the recording mode (counted or linear).*

Running any quantitative analysis on these data may be not very accurate, due to the uncertain efficiency of this method. Yet, it is possible to notice how the counted mode generally produces better results than the linear mode and that the application of the median filter significantly increases the quality of the images (*Fig. S5*).